\documentclass[twocolumn, amssymb, nobibnotes,superscriptaddress, 
aps,prmaterials]{revtex4-2}
\usepackage{graphicx}
\usepackage{xcolor}

\usepackage{hyperref}

\begin{document}

\title{Controlling the electronic interface properties of AlO$_x$/SrTiO$_3$ heterostructures}

\author{Berengar Leikert}
\affiliation{Physikalisches Institut and W\"urzburg-Dresden Cluster of 
Excellence ct.qmat, Universit\"at W\"urzburg, D-97074 W\"urzburg, Germany}
\author{Judith Gabel}
\affiliation{Physikalisches Institut and W\"urzburg-Dresden Cluster of 
Excellence ct.qmat, Universit\"at W\"urzburg, D-97074 W\"urzburg, Germany}
\affiliation{Diamond Light Source, Harwell Science and Innovation Campus, 
Didcot, OX11 0DE, UK}
\author{Matthias Schmitt}
\author{Martin St\"ubinger}
\author{Philipp Scheiderer}
\author{Louis Veyrat}
\affiliation{Physikalisches Institut and W\"urzburg-Dresden Cluster of 
Excellence ct.qmat, Universit\"at W\"urzburg, D-97074 W\"urzburg, Germany}
\author{Tien-Lin Lee}
\affiliation{Diamond Light Source, Harwell Science and Innovation Campus, 
Didcot, OX11 0DE, UK}
\author{Michael Sing}
\author{Ralph Claessen}
\affiliation{Physikalisches Institut and W\"urzburg-Dresden Cluster of 
Excellence ct.qmat, Universit\"at W\"urzburg, D-97074 W\"urzburg, Germany}
\date{\today}%

\begin{abstract}

Depositing disordered Al on top of SrTiO$_3$ is a cheap and easy way to create 
a two-dimensional electron system in the SrTiO$_3$ surface layers. To 
facilitate future device applications we passivate the heterostructure by a 
disordered LaAlO$_3$ capping layer to study the electronic properties by 
complementary x-ray photoemission spectroscopy and transport measurements on 
the very same samples. We also tune the electronic interface properties by 
adjusting the oxygen pressure during film growth.
\end{abstract}

\maketitle

\section{Introduction}

The famous high-mobility two-dimensional electron system (2DES) at the epitaxial
interface of the band insulators SrTiO$_3$ (STO) and LaAlO$_3$
(LAO)\cite{ohtomo} with its multiple intriguing electronic properties such as
superconductivity, magnetism and even the coexistence of both \cite{li} has
sparked intense interest in transition-metal oxide heterostructures.
\cite{ohtomo2, chen_high, passarello, schutz}. These materials and the LAO/STO
2DES in particular are a fascinating wellspring for novel functionalities, for
example in memresistive devices \cite{maier}. However, heterostructuring with
epitaxial LAO layers is not the only way to create a 2DES in STO. It can also be
generated, e.g., by capping with disordered LAO (d-LAO) \cite{chen_metallic} or 
by irradiation with intense synchrotron light \cite{meevasana,lenart}. A 
particular interesting route was recently reported by R\"odel \textit{et al.} 
\cite{rodel}, who use a very simple and easily upscalable approach, namely 
room-temperature evaporation of Al on a STO surface. This results in a 
disordered and partially oxidized Al film, where the oxygen is scavenged from 
the STO creating O vacancies near the surface which in turn act as $n$-dopants 
and thus give rise to the 2DES. Their \textit{in situ} angle-resolved 
photoemission spectroscopy (ARPES) data shows that just about one monolayer of 
Al suffices to create a 2DES.

The system AlO$_x$/STO has been studied also by other groups. Fu and Wagner 
\cite{fu}
evaporated 6\,nm of Al on top of STO and analyzed the Al oxidation state by
x-ray photoemission spectroscopy (XPS) as function of annealing temperature 
after film deposition. They observed partial oxidation of the Al layer even at 
room temperature and strongly enhanced oxidation upon further annealing. In 
another XPS study on multiple metal/STO interfaces Posadas \textit{et al.}
\cite{posadas} focused on the role various material parameters play for 2DES
formation, namely the workfunction difference between metal and STO as well as
the heat of formation for metal oxide versus oxygen vacancies in STO. They
confirm that the redox reaction between Al and the STO surface saturates after
about one monolayer of deposited Al, with the resulting metallicity of the STO 
surface indicated by a shoulder of reduced Ti$^{3+}$ in the Ti\,2\textit{p} 
spectrum. {Very recently, Vicente-Arche \textit{et al.} also observed a reduction of the STO surface layers upon depositing various metals, amongst them Al, in photoemission spectroscopy \cite{vicentearche2021metalsrtio3}. They also confirm that the interfacial redox reaction saturates after a certain metal thickness and observe metallic behavior in \textit{ex situ} transport measurements.}

{Several other groups} have studied quantum
transport in the AlO$_x$/STO system. For example, Sengupta \textit{et al.}
\cite{sengupta} confirmed metallicity and even studied the
superconducting regime of the 2DES by application of gate voltages and magnetic 
fields.
Wolff \textit{et al.} \cite{wolff} reported transport experiments on AlO$_x$/STO
samples grown by pulsed laser deposition (PLD) using an Al$_2$O$_3$ target for
film growth. This growth technique is, however, more complicated, less suitable
for upscaling, and may affect the formation of the AlO$_x$/STO interface in a
different way due the high kinetic energy ions in the PLD plasma plume.
Vaz \textit{et al.} \cite{vaz} even went one step further by combining the
AlO$_x$/STO interface with a NiFe layer for spin pumping experiments, thereby
paving the way for novel oxide based spintronic devices.

We note that all of
these transport experiments were performed \textit{ex situ}, i.e.,
outside the vacuum of the growth chamber. Unfortunately, so far no systematic
information has been provided how exposure to air might affect the 
pristine sample properties as probed in the aforementioned 
photoemission studies {\cite{anmerkung}}.

With this in mind and motivated by its intriguing simplicity we elaborate in
this study on the fabrication of AlO$_x$/STO by thermal Al evaporation. We
develop a reliable passivation layer for the heterostructure enabling the 
direct correlation of \textit{in situ} photoemission spectroscopy and 
\textit{ex situ} electrical transport measurements on the same device-like 
samples. By combining the results
of both techniques we demonstrate how the growth parameters can be used to
control the carrier density of the ensuing 2DES.


\begin{figure*}[ht]
	\centering
	\includegraphics[width=1\textwidth,keepaspectratio]{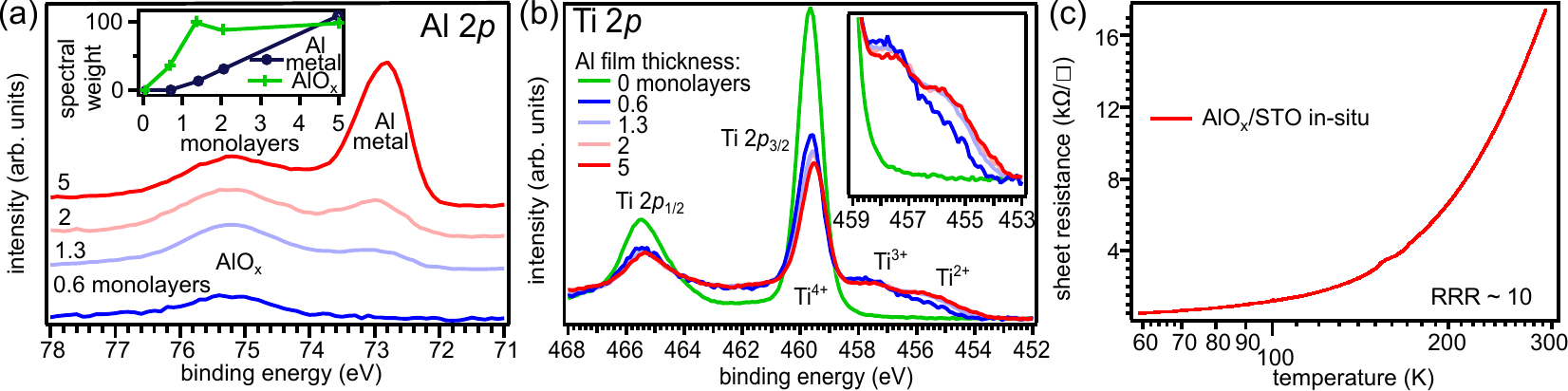}
\caption{(a) Al\,2\textit{p} spectra dependent on film thickness. The spectra
are offset for clarity and normalized to integral Ti\,3\textit{s} spectral
weight. The inset shows the spectral weight of the features as function of film
thickness. (b) Ti\,2\textit{p} spectra of the samples in (a) at the 
corresponding Al film thicknesses in contrast to the Ti\,2\textit{p} 
spectrum of bare Nb:STO. The significant, chemically shifted Ti$^{3+}$ and 
Ti$^{2+}$ features, developing upon Al deposition, indicate the formation of 
oxygen vacancies, which dope the STO with electrons. The reduction of Ti 
saturates at the
same film thickness as the Al oxidation. The spectra are normalized to integral
Ti\,2\textit{p} spectral weight. (c) Temperature dependent sheet resistance of a
sample with an Al film thickness of 5\,nm. The red curve 
corresponds to a measurement before exposure to air, i.e., it was 
recorded \textit{in situ}. It shows metallic behavior with a residual 
resistivity ratio (RRR) of about 10, while the resistance of 
the sample was beyond the measurement limit after exposure to ambient 
conditions, i.e., when measured \textit{ex situ}.
	\label{bild1}}
\end{figure*}

\section{Experimental details}\label{setup}

The STO substrates from Crystec GmbH were TiO$_2$ terminated as described
elsewhere \cite{kawasaki,koster}. Immediately before the Al growth they were
annealed inside a vacuum chamber at 500\,$^{\circ}$C in a $1\times10^{-5}$\,mbar
O$_2$ atmosphere. After the substrates had cooled down to room temperature,
aluminum was deposited by thermal evaporation from an effusion cell, in
ultrahigh vacuum (UHV) or under various oxygen background pressures. The film
thicknesses were derived by monitoring the evaporation rate with a quartz
microbalance and calibrating the flux against x-ray reflectivity data measured
for of a thick Al film grown on STO. After growth the samples were transferred
\textit{in situ} into a photoemission chamber and characterized by XPS, using a
monochromated Al K$_{\alpha}$ x-ray source and an Omicron EA-125 spectrometer.
Before the samples were removed from vacuum for the \textit{ex situ} transport
experiments, they were capped by 3\,nm of disordered LAO. This was achieved
after another \textit{in situ} transfer into a dedicated UHV chamber for
pulsed-laser deposition. The disordered LAO films were ablated from a single
crystal target at room temperature, with a laser flux of 1.3\,J\,cm$^{-2}$, a
pulse frequency of 1\,Hz, a target-substrate distance of 5\,cm and \textit{in
vacuo}. In order to establish that the capping does not alter the spectroscopic
properties of the films the passivated samples were probed by hard x-ray
photoelectron spectroscopy (HAXPES) at beamline I09 of the Diamond Light Source
(UK), using a photon energy of $h\nu=3$\,keV and a sample temperature of 50\,K.\\
Transport measurements of the passivated samples were performed in a Physical
Property Measurement System (PPMS) by Quantum Design. For these experiments the
samples were contacted by ultrasonic Al wire bonding in van der Pauw geometry. {Longitudinal and Hall resistance was measured in a temperature range between 2\,K and 300\,K and at magnetic fields up to $\pm$9T perpendicular to the sample surface.} A
small set of samples was also mounted insulating on a dedicated sample holder
and contacted before growth to enable \textit{in situ} transport experiments
for comparison.


\begin{figure*}
	\centering
	\includegraphics[width=1\textwidth,keepaspectratio]{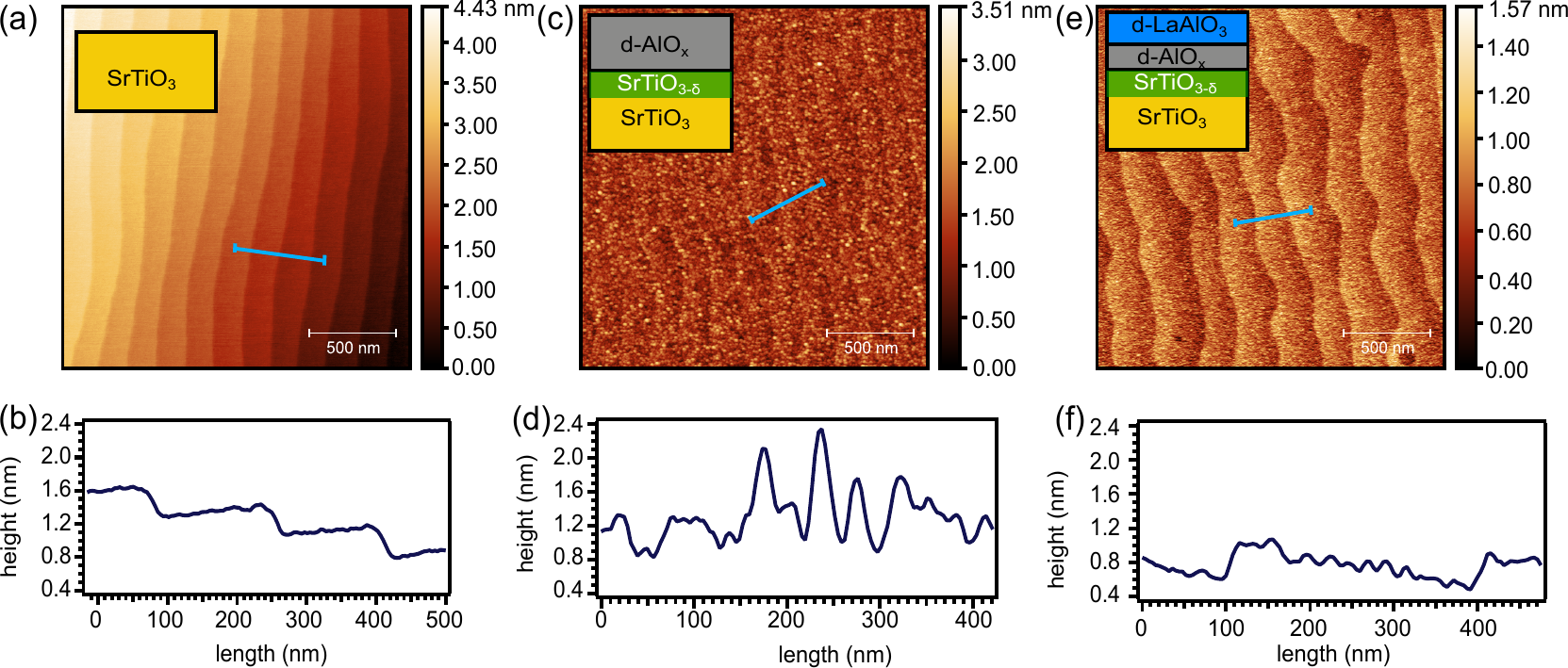}
\caption{(a) AFM image of an etched, TiO$_2$ terminated STO substrate displaying
large, flat terraces. (b) Profile along the line indicated in (a). The steps
between the terraces are 0.4\,nm high, corresponding to one STO unit cell. (c)
AFM image of a nominally 5\,nm thick, disordered AlO$_x$ film on top of a STO 
substrate.
Terraces, as imposed by the substrate, are barely seen. Rather large AlO$_x$
clusters are formed which appear as peaks with steep slopes in AFM line
profiles. An example is shown in (d) along the line indicated in (c). The peak
heights amount to over 1\,nm. (e) AFM image of 3\,nm d-LAO on top of 
nominally 2\,nm AlO$_x$ on STO [total nominal film thickness of 5\,nm as in 
(c)]. Terraces are clearly
distinguishable. The film roughness is smaller than in (c). (f) Profile along
the line in (e).
\label{AFMbild3}}
\end{figure*}

\section{Experimental results}\label{Experimental results}


\subsection{XPS analysis of pristine AlO$_x$/STO} 

We begin by calibrating our pristine AlO$_x$/STO films grown in UHV, i.e., 
without any additional oxygen supply, against the results of
previous studies in the literature, using XPS as spectroscopic characterization
tool. For this purpose we show in Fig. \ref{bild1}(a) the Al\,2\textit{p}
spectrum as function of film thickness. Starting with the lowest Al coverage
(nominally corresponding to 0.6 monolayers) we observe a single peak at 75.2\,eV
binding energy which is the value expected for fully oxidized Al \cite{nist}.
With increasing coverage an additional and chemically shifted component at
72.8\,eV can be seen to develop. It is readily identified as due to metallic Al
\cite{nist} and strongly increases in spectral weight relative to the oxide peak
upon further evaporation. This behavior is fully consistent with the
observations of R\"odel \textit{et al.} \cite{rodel}. In the absence of other
oxygen sources these spectra strongly suggest that the first Al atoms deposited
on the substrate surface become oxidized to AlO$_x$ (mostly Al$_2$O$_3$) by
scavenging oxygen from the top STO layers. Interestingly, the intensity of the
AlO$_x$ peak remains rather constant for increasing Al coverages [cf. inset Fig.
\ref{bild1}(a)], indicating that the redox process saturates already after one
monolayer while all additional Al stays in a metallic state.

Independent evidence that the oxidation of the first Al layer creates oxygen
vacancies near the AlO$_x$/STO interface is provided by the Ti\,2\textit{p}
spectrum [see Fig. \ref{bild1}(b)]. While the bare substrate displays only a
single Ti$^{4+}$ peak, we detect additional Ti$^{3+}$ and Ti$^{2+}$ signals
\cite{nist} upon Al deposition, best observed on the lower binding energy side
of the Ti\,2\textit{p}$_{3/2}$ peak. {The reduced Ti valences are a fingerprint of
the formation of single or even clustered oxygen vacancies (in the case of Ti$^{2+}$) on the STO side of the interface \cite{eom}.} The relative
intensity of these peaks with respect to that of the tetravalent, i.e., fully
oxidized, Ti species displays very little change for Al coverages above one
monolayer, again pointing to a saturation of the interfacial redox reaction. 

The oxygen vacancies thus act as \textit{n}-dopants which donate electrons to 
the STO conduction band states near the interface, thereby forming the 2DES
\cite{posadas,sing_profiling,berner,gabel_dis,scheiderer}, as is demonstrated by
the temperature-dependent resistivity curve in Fig. \ref{bild1}(c) for a 
nominally 5\,nm thick AlO$_x$ film. Note that this resistivity data 
has been 
measured \textit{in situ} by using a special set-up in our vacuum system that 
avoids exposure of the sample to atmosphere. Due to experimental 
limitations our \textit{in situ} resistivity measurements could not access the 
temperature range below 60 K. After exposing the film intentionally to ambient 
conditions, the sheet carrier resistance was beyond the measurement limit.

\subsection{Surface passivation}
\label{capping}

For a more systematic transport characterization, going down to liquid helium
temperatures and including also the measurement of the Hall resistivity, we had
to take the samples out of the vacuum in order to transfer them to our PPMS. For
such \textit{ex situ} experiments it is of paramount importance that air
exposure does not cause the created oxygen vacancies to become refilled again. 
Indeed, all our films with nominal thicknesses up to 5\,nm were found to 
become
insulating upon exposure to air. We attribute this to their higher surface
roughness compared to the bare substrate, as seen in the atomic force microscopy
(AFM) images of an STO sample before and after Al deposition in Fig.
\ref{AFMbild3}(a)-(d). While the bare substrate terraces are still atomically
flat, the AlO$_x$ film shows substantial surface corrugations on the scale of 
1\,nm or more. This implies that between thicker AlO$_x$ grains the film may 
become too thin to prevent oxygen diffusion into the STO. As a result, oxygen 
atoms refill the vacancies which eventually renders the samples insulating.

One possible solution {to} this problem could be the use of thicker Al films,
utilizing the fact that Al develops a natural oxide layer of the order of 2\,nm
in air \cite{2008} which then can act as a protecting capping. While this
approach seems to have worked well in previous studies {\cite{rodel,vicentearche2021metalsrtio3,sengupta}}, 
we found it problematic. On the one hand a passivation layer based on Al must 
not be too thin in order to suppress O diffusion to the substrate. If on the 
other hand the Al film is too thick, a metallic, i.e., conducting Al 
layer may form between the passivating surface oxide and the insulating AlO$_x$ 
phase at the interface, which in transport measurements will shunt the response 
from the 2DES in the STO. This is indeed what we observed for many of our 
thicker Al films. 

In search for a less critical passivation layer we discovered that disordered
LAO (d-LAO), grown by room-temperature PLD on top of our AlO$_x$/STO films, is a
viable option. Independent of its crystalline state (long-range ordered or
disordered) and insensitive to slight variations in stoichiometry (controlled by
the PLD growth parameters) it remains always insulating. Furthermore, d-LAO not
only prevents O-diffusion to the substrate, it also tends to stabilize the 
2DES as it is known to induce O vacancies in STO by itself
\cite{sambri}. Deposition of the protecting d-LAO layer leads to a significant
smoothening of the surface compared to the uncapped AlO$_x$ films, as is
depicted in Fig. \ref{AFMbild3}(e,f) showing AFM data of a sample with 
nominally 2\,nm AlO$_x$ capped by 3\,nm d-LAO.

In order to check the passivating power of the d-LAO capping layer we have performed 
\textit{in situ} four point resistivity measurements once before exposure to
air and once after keeping the sample two hours outside the vacuum chamber.
Figure \ref{bild2}(a) demonstrates that the temperature dependent sheet
resistance of the sample is not altered and the d-LAO film fully passivates the
heterostructure. The film stays metallic down to low temperatures with residual
resistivity ratios (RRR) of the order of 10.

\subsection{Transport experiments}

\begin{figure}
	\centering
	\includegraphics[width=1\linewidth,keepaspectratio]{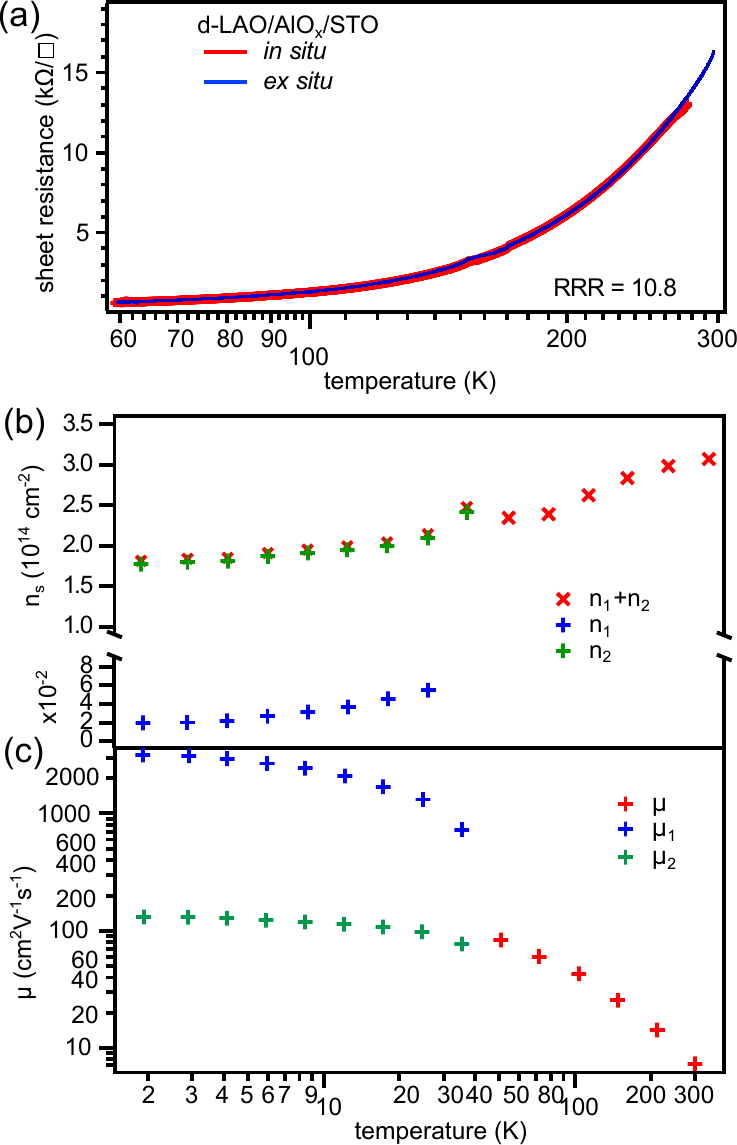}
	\caption{(a) Temperature-dependent sheet resistance of a sample grown in vacuum with a 
3\,nm d-LAO capping of nominally 2\,nm
AlO$_x$ on STO. The red curve depicts the sheet resistance of the sample
before exposure to air, i.e., when measured \textit{in situ}. Afterwards, the 
sample was taken out of the UHV cluster
and kept in air for two hours, then the measurement was repeated 
\textit{ex situ} (blue curve).
The measurements agree with each other showing that the d-LAO capping layer 
successfully
passivates the sample and enables \textit{ex situ} sample transfer. (b) {Red crosses: total sheet carrier density as function of temperature of a
d-LAO/AlO$_x$/STO sample (AlO$_x$ grown in vacuum). At low temperatures, where the Hall resistance is non-linear, the total sheet carrier density consists of the sum of $n_1$ (blue crosses) and $n_2$ (green crosses), which were extracted from a 2-band fit \cite{supplemental}. (c) Corresponding mobility data as function of temperature.}\label{bild2}}
\end{figure}

Besides the longitudinal sheet resistance we have also measured the
temperature-dependent Hall resistance of the d-LAO passivated samples. {For temperatures below 50\,K the Hall effect shows a transition from a linear behavior to a non-linear S-shape. We interpret this behavior as a sign of multiband transport, which is well-known to happen in STO based electron gases \cite{vicentearche2021metalsrtio3,joshua_universal_2012}. We fitted the nonlinear Hall effect curves using a simple 2-band model to determine the corresponding sheet carrier densities and carrier mobilities which are shown in Fig. \ref{bild2}(b) and (c) \cite{supplemental}. The model yields two electron-like carrier populations, one with large density ($n_2\approx2\,\times\,10^{14}$\,cm$^{-2}$) and low mobility ($\mu_2\approx140\,$cm$^{2}$V$^{-1}$s$^{-1}$) and one with much smaller density ($n_1\approx1\,\times\,10^{12}$\,cm$^{-2}$)) and a significantly larger mobility ($\mu_1\approx2000\,$cm$^2$V$^{-1}$s$^{-1}$).} These data also reveal that a sizable fraction of the
mobile charge carriers get frozen out at low temperatures, which is also
observed in other STO based systems where doping is dominated by the formation
of oxygen vacancies \cite{wolff,chen_metallic,trier, sengupta}. Interestingly,
the measured sheet carrier densities are at all temperatures higher than in the
AlO$_x$/STO heterostructures of Sengupta \textit{et al.} \cite{sengupta} {and the PLD-grown AlO$_x$/STO heterostructures of Wolff \textit{et al.}
\cite{wolff}, likely
due to significantly higher oxygen vacancy concentrations in our samples.}

While the carrier mobility, on the other hand, is limited at high 
temperatures by phonon scattering, it increases at lower temperatures due to
phonon freeze-out and eventually {the mobility of the majority carriers} saturates around 140 cm$^2$V$^{-1}$s$^{-1}$ at 
liquid helium temperatures where the carrier mobility is largely determined by 
impurity scattering \cite{raghavan}. The order of magnitude 
of the low
temperature {mobility of the majority carriers} is comparable to the other aforementioned oxygen
vacancy governed STO systems \cite{wolff,chen_metallic,trier} but lower than in
the samples of Sengupta \textit{et al.} \cite{sengupta}. This is again
attributed to a higher vacancy concentration in our samples, highlighting their
dual role as dopants and as scattering centers.

In order to exclude that the measured conductivity stems from metallic Al in the
film, a control experiment was conducted in which the same film was deposited on
DyScO$_3$ (DSO), another wide-gap transition-metal perovskite oxide insulator.
In contrast to STO, in DSO the electronic charges introduced by doping with
oxygen vacancies remain completely trapped, thereby preventing the formation of
a metallic 2DES \cite{yuan,velickov}. {Thus, conduction in AlO$_x$/DSO would be an indication of remaining metallic Al.} Indeed, all our AlO$_x$/DSO
heterostructures grown under identical conditions as the AlO$_x$/STO films
stayed insulating, with resistances beyond the measurement limit of the PPMS. {As metallic Al is even less likely to form in AlO$_x$/STO, where the redox reaction is stronger, than in AlO$_x$/DSO, this control experiment strongly goes in favor of a totally oxidized and insulating Al layer.}


 \begin{figure*}
	\centering
	\includegraphics[width=1\textwidth,keepaspectratio]{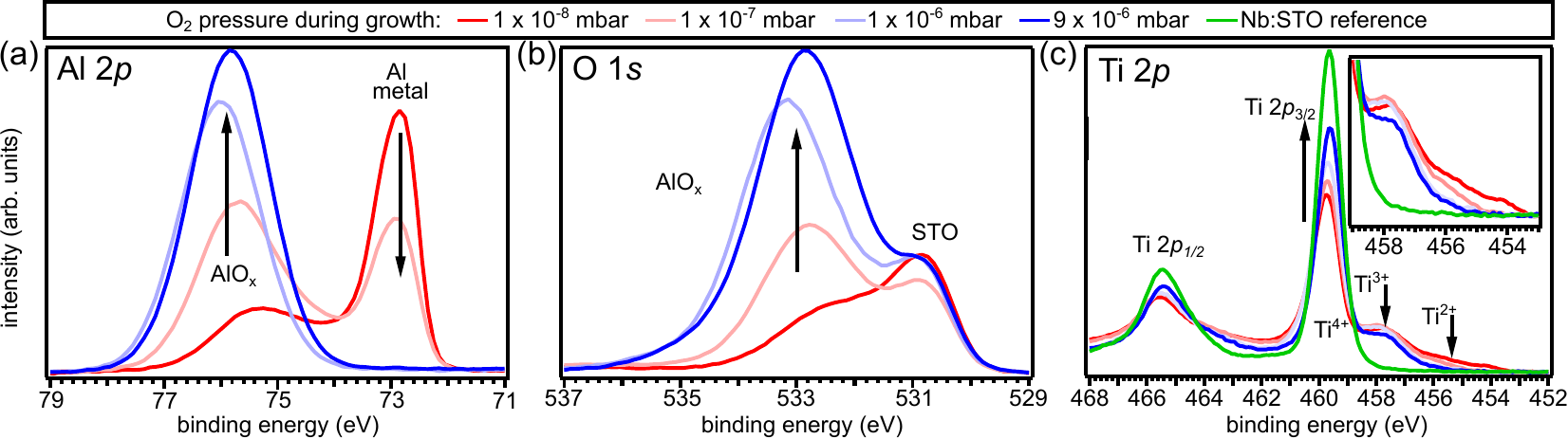}
\caption{XPS core level spectra dependent on the O$_2$ pressure during Al
deposition. (a) Al\,2\textit{p} spectra. With higher oxygen background pressure
the metallic Al component shrinks while the AlO$_x$ component grows. The
Al\,2\textit{p} spectra are normalized to the Ti\,3\textit{s} spectral weight.
(b) O\,1\textit{s} spectra of the samples in (a) normalized to the
Ti\,2\textit{p} spectral weight. (c) Ti\,2\textit{p} spectra of the same
samples, normalized to integral Ti\,2\textit{p} spectral weight. Besides the
Ti$^{4+}$ component of stochiometric STO also a chemically shifted Ti$^{3+}$
component, indicating the existence of oxygen vacancies, is detected. For 
samples
grown in low oxygen pressures, also a sizable Ti$^{2+}$ component is observed,
suggesting even stronger reduction of the STO surface. With higher oxygen growth
pressures the Ti$^{2+}$ component vanishes and the Ti$^{3+}$ component shrinks,
while the Ti$^{4+}$ component grows, showing that less oxygen vacancies are
created. \label{bild3}}
\end{figure*}


\begin{figure}
	\centering
	\includegraphics[width=1\linewidth,keepaspectratio]{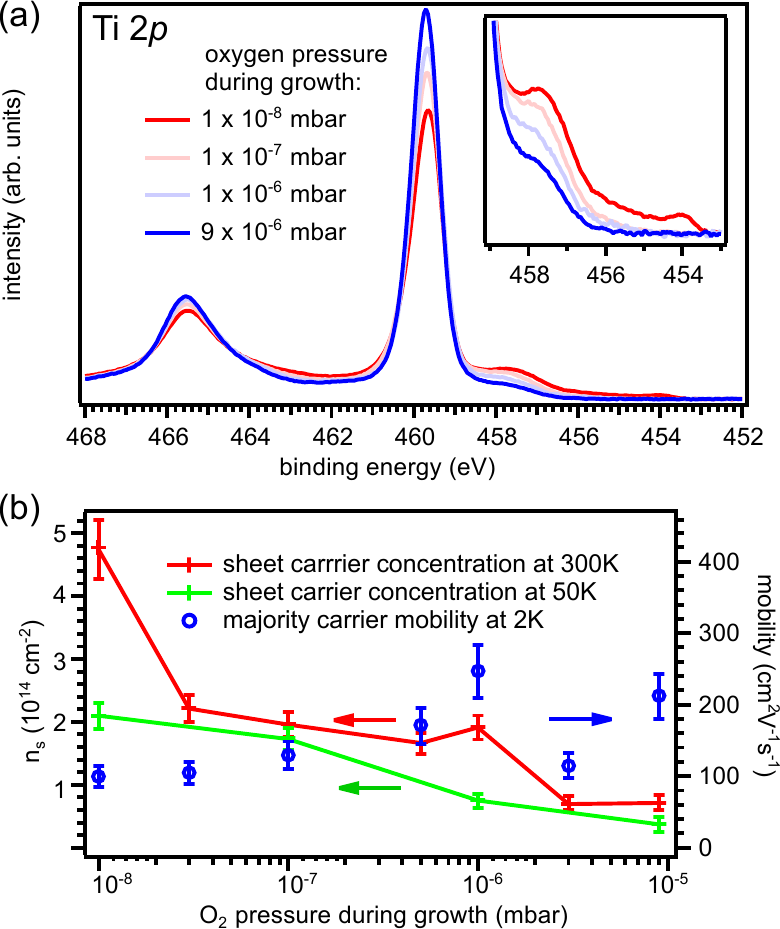}
\caption{(a) Hard x-ray photoemission Ti\,2\textit{p} spectra ($h\nu=3$\,keV) of
\textit{ex situ} d-LAO/AlO$_x$/STO samples where the Al was deposited at various
O$_2$ pressures. Inset: Zoom-in on the low binding energy side of the
Ti\,2\textit{p}$_{3/2}$ peak. (b) Left axis, red crosses: Sheet carrier
concentration at 300\,K dependent on O$_2$ growth pressure. The line is a guide
to the eye. Upon increasing the oxygen pressure during growth, the sheet carrier
density decreases. Left axis, green crosses: Sheet carrier density at 
50\,K. Right axis, blue circles: {Mobility of the majority carriers} at 2\,K dependent on O$_2$
pressure during growth. It displays no systematic behavior.\label{bild4}}
\end{figure}

\subsection{Growth control of the interface properties}

For device applications it is desirable that the electronic interface properties
can be tuned. This is achieved by varying the growth parameters, in particular
the O$_2$ background pressure applied during Al deposition. 
For a systematic study we have grown a series of
Al films with 2\,nm nominal thickness in oxygen background pressures between
$1\times10^{-8}$\,mbar and $1\times10^{-5}$\,mbar. These samples were then
characterized by \textit{in situ} XPS and \textit{ex situ} transport
measurements. Please note that all photoemission spectra were taken before
capping with d-LAO, if not stated otherwise.

We begin our analysis with a survey of the Al\,2\textit{p} spectra depicted in
Fig. \ref{bild3}(a). For comparison of their relative intensities
they have been normalized to the integral spectral weight of the Ti\,3\textit{s}
substrate core level (not shown). Note, however, that an error is involved 
here since due to oxidation, i.e., the incorporation of oxygen, the same amount 
of deposited Al will result in a larger film thickness and hence stronger 
damping of the substrate signals. Thus, the integrated intensity of the 
spectra in fact is increasingly overestimated with increasing oxygen pressure. 
The qualitative trends, however, are not affected by this normalization as will 
become clear from the following discussion of the spectra. For films grown at 
low oxygen pressures the
spectra resemble the red spectrum in Fig. \ref{bild1}(a), consisting of a
metallic Al signal and a weaker oxide component at higher binding energy. For
higher oxygen background pressures, the metallic Al peak is observed to decrease
in intensity, while the AlO$_x$ peak grows until only the latter is detected,
demonstrating that the Al gets considerably stronger oxidized if deposited under
a higher oxygen pressure.

This conclusion is confirmed by the behavior of the O\,1\textit{s} spectra 
in Fig. \ref{bild3}(b). These spectra are normalized to the integrated 
intensity of the Ti\,2\textit{p} substrate signal, chosen for its energetic 
vicinity [cf. Fig. \ref{bild3}(c)]. As above, there is a certain quantitative 
error concomitant with this normalization that, however, does not affect the 
qualitative interpretation of the data. The peak at 531\,eV is attributed to 
the oxygen in the STO substrate and stays approximately constant in intensity 
throughout the pressure series. For the lowest O$_2$ pressure an additional 
shoulder at 533\,eV corresponding to oxygen in AlO$_x$ is detected. With 
increasing 
oxygen background pressure it evolves into a pronounced peak that surpasses the 
substrate peak. Note that with the chosen normalization the spectral weight of 
the O 1s signal is a direct measure of the amount of oxygen in film and 
substrate, respectively. The strong increase of the film component consequently 
signals that the increasing O$_2$ growth pressure leads to stronger oxygen 
incorporation into the AlO$_x$ films.

So far we have seen that Al deposited on STO reacts both with oxygen diffusing
out of the substrate as well as with O$_2$ supplied by the gas phase to form
AlO$_x$. This raises the obvious question, if the latter route has an effect on
the former, thereby influencing the generation of oxygen vacancies in the STO.
This can be answered by XPS on the Ti\,2\textit{p} core level, as shown in Fig.
\ref{bild3}(c). At low O$_2$ pressure, a huge Ti$^{3+}$ shoulder and even a
Ti$^{2+}$ feature are observed in addition to the Ti$^{4+}$ peak corresponding
to stoichiometric STO. At higher oxygen background pressure the Ti$^{3+}$ as
well as the Ti$^{2+}$ shoulders decrease in intensity, with the latter becoming even
completely suppressed at the highest oxygen background pressure [s. inset of 
Fig.~\ref{bild3}(c)]. We conclude that 
the strong oxidation resulting from evaporating Al in an oxygen atmosphere strongly 
reduces the scavenging of oxygen from STO, i.e., it reduces the oxygen 
vacancy 
concentration and thereby the \textit{n}-doping of the interface layers.

Before we test this conclusion further by \textit{ex situ} transport
experiments on capped samples, we first establish that the spectroscopic
information discussed above is not altered by the application of a passivating
d-LAO capping. For this purpose we have employed \textit{hard} x-ray
photoelectron spectroscopy (HAXPES) at a photon energy of $h\nu=3$\,keV, making
use of the considerably enhanced probing depth compared to conventional XPS.
This allows us to acquire a strong photoemisson signal of the STO layers of
passivated samples despite a total (nominal) film thickness of 5\,nm (AlO$_x$ 
and d-LAO).
Figure \ref{bild4}(a) shows the Ti\,2\textit{p} core level spectra of a set of
capped samples grown at different oxygen background pressures. Similar to 
the \textit{in situ} XPS data on uncapped samples in Fig. \ref{bild3}(c) 
Ti$^{2+}$ and
Ti$^{3+}$ features are clearly discernible in these spectra and become
increasingly reduced in intensity for higher O$_2$ growth pressures.

Finally, our \textit{ex situ} transport measurements confirm the metallic
character of all samples regardless of the oxygen background pressure applied
during growth. Figure \ref{bild4}(b) compiles the transport data for 
all samples, specifically the carrier density extracted from the Hall effect 
and the low-temperature mobility. The red curve depicts the sheet carrier 
density at 
300\,K, the same temperature at which the XPS experiments were performed, 
while the green curve is sheet carrier density at 50\,K, where the HAXPES spectra were taken. 
The consistently lower carrier concentration at lower temperatures is due to 
the carrier freeze out already seen in Fig.
\ref{bild2}(b). The sheet carrier density in both curves confirms the
qualitative observations from the photoemission experiments in Fig. \ref{bild3}
and Fig. \ref{bild4}(a), namely a clear reduction with increasing 
O$_2$ growth pressure. It should be noted here that the metallic Al seen in the 
XPS spectra of Fig.~\ref{bild4}(a) for samples grown at the two lowest O$_2$ 
pressures apparently does not form a percolative path shunting the conducting 
2DES because otherwise the charge carrier concentrations are expected to be 
much higher.

In contrast to the carrier concentration we do not observe any systematic
dependence of the low-temperature mobility on oxygen background pressure [blue
symbols in Fig.~\ref{bild4}(b)]. This may either result from impurity scattering 
other than by the oxygen vacancies or indicate that the effective scattering 
rate is not only controlled by the concentration but also by the spatial depth 
profile of the vacancies. We will return to this point in the following 
section.

\section{Discussion}\label{discussion}

In the first place, our combined photoemission and transport data clearly 
confirms that the 2DES
formation in AlO$_x$/STO results from a redox process between the Al film and the
substrate, generating $n$-doping oxygen vacancies in the STO. This mechanism is different
from the generic electronic reconstruction scenario discussed for
all-crystalline oxide heterostructures with a polar discontinuity like 
LAO/STO \cite{ohtomo} or $\gamma$-Al$_2$O$_3$/STO
\cite{chen_high}. Note, however, that oxygen vacancy formation can also occur in
those systems, substantially contributing to the carrier concentration of their 2DES
and making their interface properties sensitive to the specific oxygen
exposure during and even after epitaxial growth \cite{gabel_dis,schuetz2017}.
This is exactly what we see here as well. The simplicity of the deposition
process in the case of our AlO$_x$ films provides us with the unique opportunity
to control the working point of the redox reaction by supplying additional
oxygen in the gas phase. This opens a second channel for aluminum oxidation,
resulting in reduced formation of oxygen vacancies and hence a smaller sheet
carrier density. Whether this effect is due to enhanced oxidation of the Al
atoms during film growth (before or after deposition), thereby suppressing the
scavenging of oxygen from the substrate, or whether the
background pressure leads to oxygen diffusion into the substrate and direct
filling of the vacancies, cannot be decided on the data at hand. However, the fact
that all of our uncapped AlO$_x$/STO samples turned insulating upon 
exposure 
to ambient atmosphere, seems to indicate that the latter process is at least part of the story.

While on the one hand the O$_2$ growth pressure can thus be viewed as a useful
control parameter for the interface properties, the extreme air sensitivity on
the other hand would seem to prevent any practical \textit{ex situ} use of the
AlO$_x$/STO heterostructures, unless they are protected by a suitable
passivation layer. One possible candidate for such a protecting capping is the
native oxide of Al itself, as was used by Sengupta \textit{et al.} {as well as by Vicente-Arche \textit{et al.}} for their
\textit{ex situ} transport experiments \cite{vicentearche2021metalsrtio3,sengupta}. For this purpose they
simply exposed their thick Al films to air to create an oxygen diffusion
barrier. However, in contrast to our study, {the Al-films of Sengupta \textit{et al.}} were grown at
elevated substrate temperatures ($100 \dots 200^{\circ}$C), resulting in a
smoother Al surface and hence better defined oxide capping. The choice of
room-temperature growth in our case was motivated by trying to suppress the
leveling effect of temperature-enhanced oxygen diffusion when studying the impact of oxygen
background pressure during growth. The drawback of this approach is our larger
film roughness (at same nominal thickness as in Ref.~\onlinecite{sengupta}),
which upon air exposure prevents the formation of a sufficiently homogeneous
native oxide as diffusion barrier but rather leads to a complete quenching of the 2DES.
Interestingly, the films of Sengupta \textit{et al.} display a significantly
lower sheet carrier density compared to our heterostructures,
indicating that the exposure to air also in their samples causes a reduction 
of oxygen vacancies in the STO substrate, though not quite as dramatic as in 
our case. Unfortunately, they were not able to perform \textit{in situ} 
transport measurements in order to assess the changes before and after air 
oxidation.

As demonstrated in Section \ref{capping}, a thin film of disordered LAO is an
excellent alternative as reliable oxygen diffusion barrier, protecting our
heterostructures and their electronic interface properties from unwanted
oxidation when taken \textit{ex situ}. This is essential for reproducible device
fabrication and operation. Another important aspect of d-LAO is the fact that
\textit{room-temperature} growth suffices to achieve a protecting and
simultaneously electrically insulating capping, being an indispensable
prerequisite for our systematic study of the role of oxygen growth pressure. Any
capping material that would require higher growth temperatures will unavoidably
alter the amount and distribution of oxygen and oxygen vacancies in the system
in an uncontrolled way. As a consequence, this would have prevented a meaningful
direct correlation between our \textit{in situ} photoemission and \textit{ex
situ} transport data.

Other possible routes towards a surface passivation of AlO$_x$/STO
heterostructures, including Al$_2$O$_3$ deposition by atomic layer deposition or
sputtering, are viable as long as the capping layer grows flat, blocks oxygen
diffusion and is thick enough. We note in passing that a recent report on
spin-charge conversion based on the AlO$_x$/STO heterostructure successfully
used a \textit{functional} capping, consisting of a 20\,nm ferromagnetic NiFe
layer covered in turn by an additional, sputtered AlO$_x$ film \cite{vaz}.

We finally address the apparent lack of any systematic variation of the
low-temperature mobility in our AlO$_x$/STO films with oxygen growth pressure.
This observation seems surprising in view of the fact that in the studied O$_2$ pressure range
the carrier density changes monotonically by more than a factor of 5 (at
$T=50$K, factor of 8 at 300 K). Naively one
would expect the dual role of the oxygen vacancies in the STO substrate --
controlling the sheet carrier density by $n$-doping as well as acting as
scattering centers -- to cause an anticorrelation between carrier concentration
and mobility. However, this picture implicitly assumes that carriers and oxygen
vacancies share the same vertical depth profile in the heterostructure, which in
fact does not have to be the case. For example, it has recently been found that
the several orders of magnitude larger mobility of the 2DES in 
$\gamma$-Al$_2$O$_3$/STO with
respect to that of LAO/STO results from a different density profile of the
oxygen vacancies related to the specific atomic micro structure of the two
interfaces \cite{schuetz2017,christensen}. While the vacancies are tightly bound to the
spinel-perovskite boundary of the former, they spread further out from the interface
in the latter all-perovskite system. This results in a spatial
separation of conduction electrons and scatterers in $\gamma$-Al$_2$O$_3$/STO
and thus its higher mobility as opposed to LAO/STO.

Against this background it seems conceivable that the oxygen vacancy
distribution in our AlO$_x$/STO system is also affected by the structural
details of the interface and, in particular, by its sensitivity to the varying
oxidation conditions during Al deposition, potentially leading to a complex
interplay of the different depth profiles of carriers and vacancies. The
situation is further complicated by the fact that in contrast to the
above-mentioned epitaxial interfaces the AlO$_x$ films are disordered or
nanocrystalline at best. For a better understanding of the transport properties
it would nonetheless be desirable to determine the respective (laterally
averaged) depth profiles. In principle, this information can be achieved by
photoemission of the Ti\,3\textit{d}-derived features in the STO-band gap, namely
the metallic quasiparticle peak at the Fermi energy and the so-called in-gap
peak at approx.~1.5 eV \cite{berner,gabel_dis}. While the quasiparticle peak is 
directly related to the
mobile charge carriers, the in-gap peak probes Ti\,3\textit{d} electrons locally
trapped by oxygen vacancies. Measuring their angle-dependencies provides the
depth profiles of carriers and oxygen vacancies, respectively, as has
successfully been demonstrated for other STO-based heterostructures
\cite{cancellieri2013,schuetz2017}. The detection of Ti\,3\textit{d}-derived 
interface
states requires resonant photoemission at the Ti $L$ edge which unfortunately 
has
not been available for the present study but is planned for forthcoming
experiments.

\section{Conclusion}\label{summary}
Using combined x-ray photoemission spectroscopy and transport measurements on 
identical samples---both \textit{in} and, in particular, \textit{ex situ}, the 
latter being enabled by the use of suitable, passivating disordered LaAlO$_3$ 
capping layers---we have demonstrated the tunability of the electronic 
interface properties of AlO$_x$/SrTiO$_3$ heterostructures. Upon depositing Al 
on the SrTiO$_3$ surface, oxygen is incorporated in the Al film leading to 
oxygen vacancies in the SrTiO$_3$ surface region that in turn release electrons 
to form a two-dimensional electron system. Its charge carrier concentration can 
be finely adjusted by the oxygen growth pressure, while its mobility seems to 
subtly depend on the detailed redox reactions, kinetics, and thermodynamics, 
taking place during film deposition, and thus cannot be easily controlled. Our 
results may be relevant for the development of device-like architectures that 
rely on the low-dimensional electron systems known to be inducible in SrTiO$_3$.

\begin{acknowledgments}

The authors thank Michael Zapf for helpful discussions. We acknowledge Diamond
Light Source for time on beamline I09 under proposal SI15856. The authors also
thank D. McCue and D. Duncan for technical support at beamline I09.

The authors are grateful for funding support from the Deutsche 
Forschungsgemeinschaft (DFG, German Research Foundation) under Germany's
Excellence Strategy through the W\"urzburg-Dresden Cluster of Excellence on 
Complexity and Topology in Quantum 
Matter ct.qmat (EXC 2147, Project ID 390858490) as well as through the 
Collaborative Research Center SFB 1170 ToCoTronics (Project ID 258499086).

\end{acknowledgments}

%


\end{document}